\documentstyle[sprocl]{article}
\begin{document}
\input epsf
\def\Se{S_{\hbox{\scriptsize eff}}}
\def\ds{\displaystyle}

\def\GS{G_{\mbox{\scriptsize S}}}
\def\GV{G_{\mbox{\scriptsize V}}}
\def\LC{\lambda_{\mbox{\tiny c}}}
\def\MF{m_{\mbox{\tiny f}}}
\def\CFF#1#2{\centerline{\Black\epsfxsize=#1\epsfbox{#2}}}

\font\Figfont=cmr8
\def\Black{}
\let\Blue\Black
\let\Cyan\Black
\let\Red\Black
\let\Green\Black
\let\Magenta\Black
\let\White\Black
\let\Yellow\Black
\let\Violet\Black
\null\vskip -15mm
\hbox to \hsize{\hfil KANAZAWA-97-09}

\title{
NON-PERTURBATIVE RENORMALIZATION GROUP APPROACH TO
DYNAMICAL CHIRAL SYMMETRY BREAKING IN GAUGE THEORIES
\footnote{Invited talk at International Workshop on
Perspectives of Strong Coupling Gauge Theories (SCGT96), 
Nagoya, 1996.
This talk is based on the work in collaboration with 
K.~Morikawa, K.~Shimizu, W.~Souma, J.-I.~Sumi, M.~Taniguchi, 
H.~Terao and M.~Tomoyose.}
}

\author{ Ken-Ichi AOKI }

\address{Department of Physics, Kanazawa University\\
Kakuma-machi, KANAZAWA 920-11\\
JAPAN}

\maketitle\abstracts{
We analyze the dynamical chiral symmetry breaking in gauge 
theories solely by the non-perturbative renormalization group, 
without recourse to the Schwinger-Dyson method.
First, we briefly review the basic notions and formulation,
and clarify its great feature that it gives
a systematic approximation scheme without any divergent series nor
serious gauge dependence, compared to
other perturbative and non-perturbative methods.
Then we apply this new method to QED and 
QCD to find that our lowest level approximation improves the ladder
Schwinger-Dyson equation results in a gauge independent way. 
We get the chiral phase structures 
and the critical exponents in the fixed gauge coupling analysis, 
and also calculate the chiral condensates in case of running QCD.
}
  
\section{Introduction}

Our aim is to attack the dynamical chiral symmetry 
breaking solely by the non-perturbative renormalization group
(NPRG). The ladder Schwinger-Dyson (SD) equation
has been the best analytic tool 
to investigate it,\cite{SDmain,SDfourfermi} but is 
unsatisfactory due to the strong gauge dependence,\cite{SDdifficulty}
and difficulty of going beyond the ladder,\cite{SDbeyond} while the
improved ladder approximation with the running gauge coupling constant
gives good results in QCD even
quantitatively.\cite{Higashijima,SDimproved}
However the improved ladder is just a model, 
having no firm theoretical base, thus there has been no way to 
further improve it.

As everybody knows, NPRG must best fit the issue. We first 
define a renormalization group equation
for some appropriate sub-theory space, then we get
flows, critical surfaces and fixed points, and we can
calculate critical exponents or anomalous dimensions etc.
However nobody had known a good way of approximation to perform
this strategy. 
Recently we found a good approximation, which is a systematic
and consistent method, contains the ladder and even the improved ladder
SD results, 
thus we can go beyond the ladder, can improve the improved ladder, in 
a gauge independent way. It may be applicable to wide
range of models and phenomena including supersymmetry, dynamical
gauge symmetry breaking, topological effects, etc.

In a battle field 
map of the renormalization group approaches (Fig.\ref{fig-map}),
NPRG is just in the mid of the
field, between the perturbative and the non-perturbative 
head quarters. It shares good features of both sides.
The lattice simulation is in fact the most prominent
in the non-perturbative regime, but it suffers serious
problems of chiral fermions, and that it has exhausted the 
computer resources on the earth, although recently
improved actions enhanced it even thousand times.
It is not feasible to simulate systems with large
hierarchies. Its great success in QCD is due to the fact that
the confinement transition can be described within a relatively
 short range of energy scales.

\refstepcounter{figure}
\label{fig-map}
\refstepcounter{figure}
\label{fig-map-a}
\vskip5mm
\hbox to \hsize{\hfil\epsfxsize=60mm\epsfbox{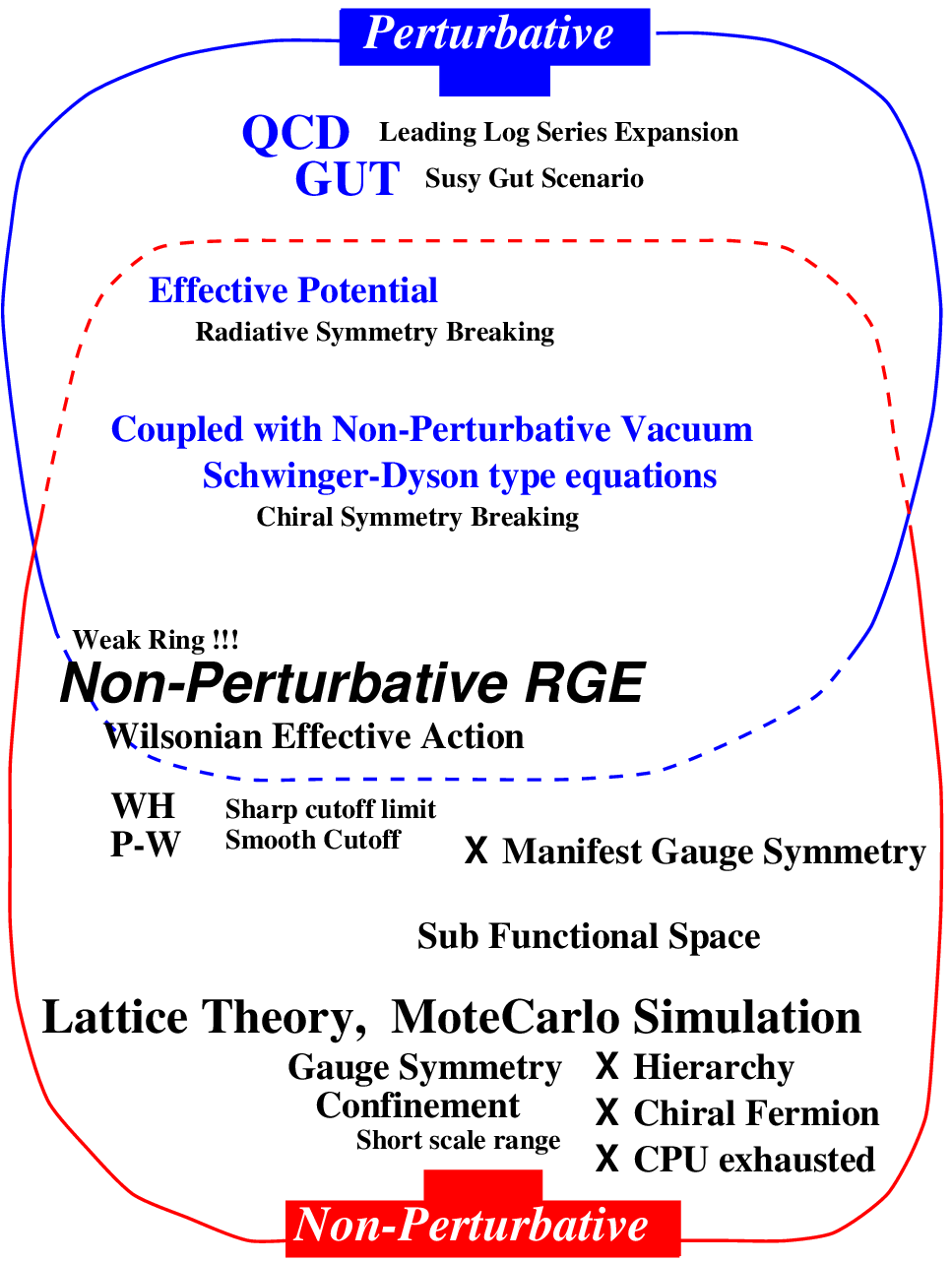}
\hfil\epsfxsize=50mm\epsfbox{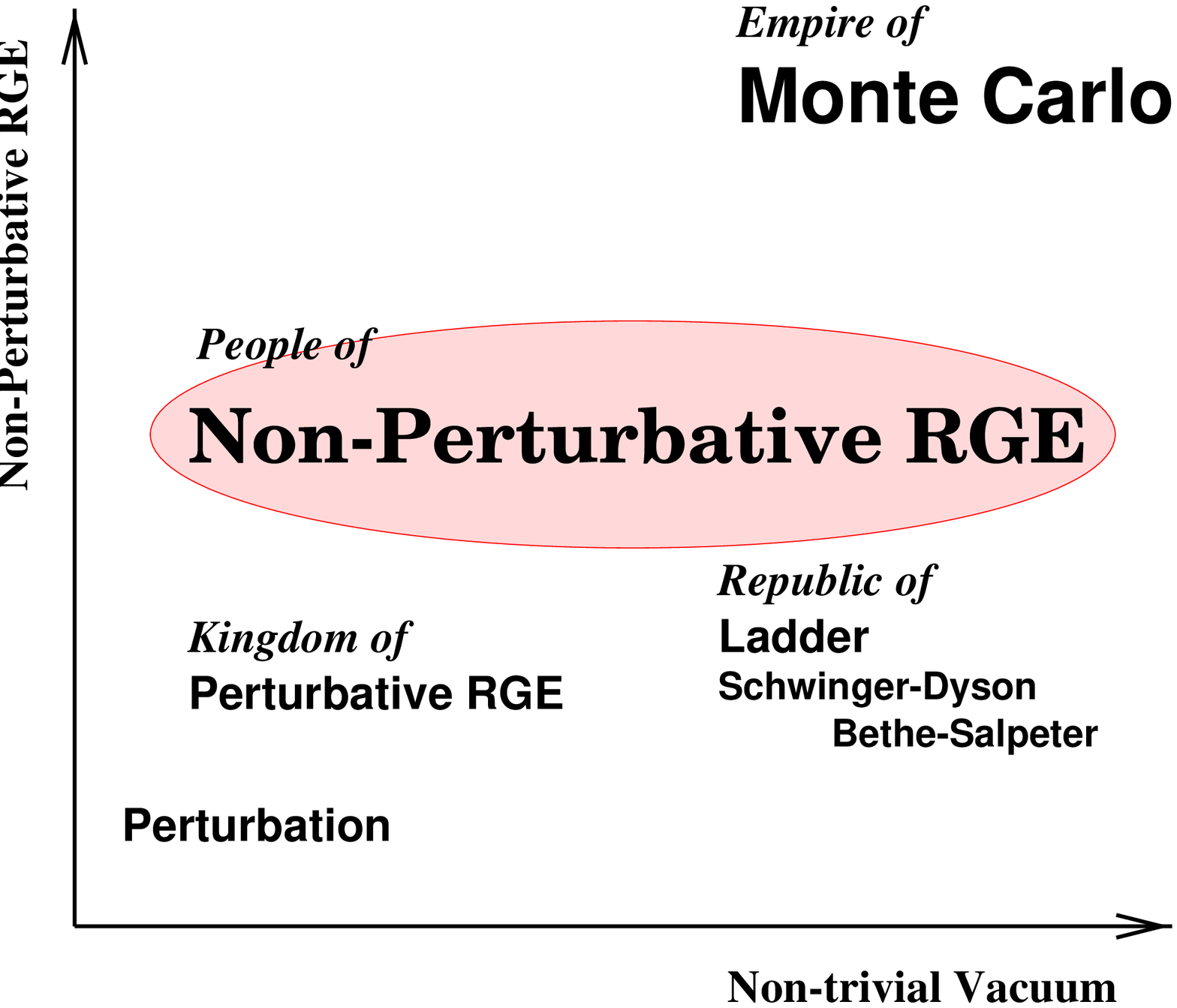}\hfil}
\hbox to \hsize{\Figfont\hfil\hbox to 60mm{\hfil Fig.\ref{fig-map}\hfil}
\hfil\hbox to 50mm{\hfil Fig.\ref{fig-map-a}\hfil}
\hfil}

On the other hand, in NPRG, there are no difficulty of chiral fermions 
nor large hierarchies. The most serious
problem in NPRG is that it cannot respect the manifest gauge symmetry
due to the introduction of the momentum cutoff.
However this is not an essential difficulty, but just a 
practical issue. One should remember that the renormalization 
calculation in gauge theories could be done even with the momentum 
cutoff regularization, as was done for QED before
the gauge invariant regularization was invented.

Figure \ref{fig-map-a} shows a more personal view of the field. 
We, the Wilsonian people, is surrounded by other old countries.
In other words we share all beauties of these countries
simultaneously. We are now lonely attacking the weakest
neighbor of the republic of ladder also known as Nagoya, and 
finally will merge it, we hope.


\section{Wilsonian renormalization group}

In short, NPRG is a method to evaluate
integral by solving a corresponding differential equation,
$$ F(t) \equiv \int^t_0 f(x) dx\ \  
\Longrightarrow\ \  {\partial F(t) \over \partial t} = f(t),\ \ 
F(0)=0, $$
which is nothing but a definition of integral, though.
We evaluate a path integral by solving a functional
differential equation which is called NPRG 
equation.\cite{WILSON,WH,Polchinski,WETTERICH}
The essential ingredient is the Wilsonian effective action 
$\Se$ defined by 
$$
\int{\cal D}\phi\exp\left(- S[\phi]\right)
=\int{\cal D}\phi_>{\cal D}\phi_<\exp\left(- S[\phi_>, \phi_<]\right)
\equiv\int{\cal D}\phi_<\exp\left(- \Se[\phi_<]\right) ,
$$
$$
\exp\left(-\Se[\phi_<]\right)
=\int{\cal D}\phi_>\exp\left(- S[\phi_>,\phi_<]\right)\ .
$$
The high frequency modes ($\phi_>$) are integrated out, 
and the effective action
for the low frequency modes ($\phi_<$) is obtained. This is equivalent to the 
so-called block spin transformation, and it physically describes 
the coarse graining or the transition from {\sl micro} to {\sl macro}.

Now, we define the flow of $\Se$ by 
continuously decreasing the cutoff $\Lambda$ which is the boundary between
the high and low frequency modes.
The effective action at $\Lambda$, $\Se[\Lambda]$, is obtained after 
the high frequency modes ($>\Lambda$) have been integrated out:
$
Z=\int^\Lambda{\cal D}\phi\exp\left(- S[\phi; \Lambda]\right)\ ,
$
where the measure ${\cal D}\phi$ contains low modes only.
We evaluate the derivative of $\Se[\Lambda]$, 
$\ds{\partial \over\partial\Lambda}\Se[\phi; \Lambda]$, where
the fictitious time parameter $t$ is introduced to control the
cutoff: $\Lambda(t) = \exp(-t)\Lambda, 
\ \delta\Lambda = \Lambda\delta t$.
Suppose we lower the cutoff by $\delta\Lambda$, and evaluate the
change of the effective action $\delta\Se$, which comes from the
shell mode ($[\Lambda - \delta\Lambda, \Lambda]$) path integral.
We expand $\Se$ with respect to the shell mode field
$\phi_{\mbox{s}}$,
\begin{eqnarray*}
&&\exp\{-\Se[\phi_{<};\Lambda-\delta\Lambda]\}
=\int{\cal D}{\Red\phi_{\mbox{s}}}
\exp\{-\Se[\phi_{<}+{\Red\phi_{\mbox{s}}};\Lambda]\}\\
&=&\exp(-\Se[\phi_{<};\Lambda])
 \int {\cal D}{\Red\phi_{\mbox{s}}}
\exp\left\{ 
-\int'_p
\frac{\delta \Se}{\delta\phi_p}{\Red\phi_{\mbox{s}}}(p)
\right.
\\
& &-\frac{1}{2}
\left.
\int'_p \int'_q
\frac{\delta^2 \Se}{\delta\phi_p \delta\phi_q}
{\Red\phi_{\mbox{s}}}(p)
{\Red\phi_{\mbox{s}}}(q)
+\cdot\cdot\cdot \right\}\\
&=&\exp\{-\Se[\phi_{<};\Lambda]\}
\int{\cal D}{\Red\phi_{\mbox{s}}}\\
& &
\exp\left\{
-\int'_p \left(\frac{\delta \Se}{\delta \phi_p}{\Red\phi_{\mbox{s}}}(p)
+\frac{1}{2}{\Red\phi_{\mbox{s}}}(p)
\frac{\delta^2 \Se}{\delta\phi_p \delta\phi_{-p}}
{\Red\phi_{\mbox{s}}}(-p)\right)
+O(\delta \Lambda^2)
\right\}\ ,\\
\end{eqnarray*}
\vskip-6mm
\noindent
where the momentum integration is carried over the shell momentum
denoted by the prime.
Only the first order term in $\delta\Lambda$ must be kept to evaluate
the derivative. The terms with more than two shell modes do not 
contribute to the first order, since they must be accompanied by 
at least two loop integral of the shell momentum, thus leaving 
$\mbox{O}(\delta\Lambda^2)$.
Then the shell mode path integral can be exactly carried out, which 
is just the Gaussian integral, resulting
$$
\delta \Se
=\frac{1}{2}\int'dp
\left\{
-\hbox{Tr}\ln
\left(
\frac{\delta^2 \Se}{\delta\phi_p\delta\phi_{-p}}
\right)
+
\frac{\delta \Se}{\delta\phi_p}
\left(
\frac{\delta^2 \Se}{\delta\phi_p\delta\phi_{-p}}
\right)^{-1}
\frac{\delta \Se}{\delta\phi_{-p}}
\right\}\ .
$$
Graphically these contributions are
represented by the following diagrams, {\sl ring} and {\sl dumbbell}, and
its physical reasoning is clear.
The shell mode path integral produces
all the quantum corrections due to the shell modes, with external 
fields of the low modes. Note that the tree and the one loop diagrams 
complete the
derivative, and no external momenta may flow in the loop except for the
dumbbell endpoint vertices. 
\vskip1mm
\centerline{\epsfxsize=10cm\epsfbox{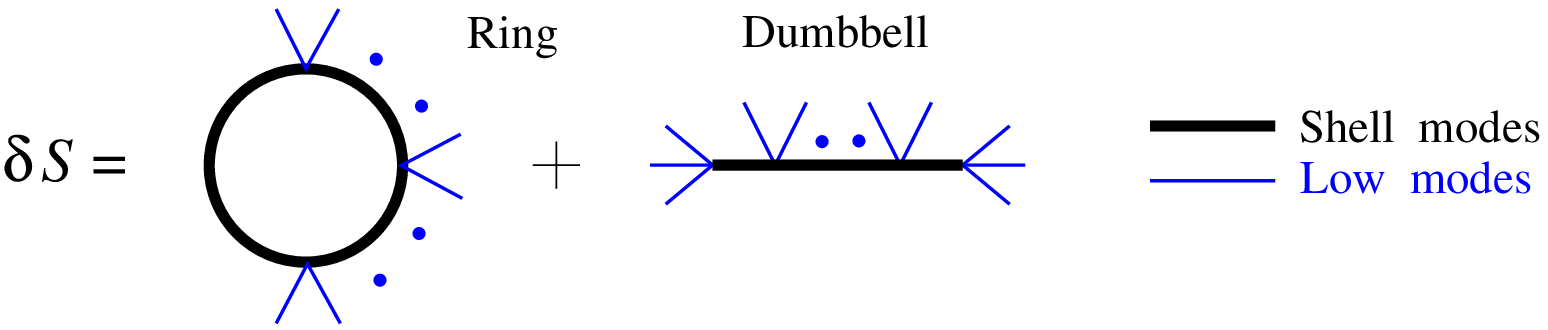}}

To make NPRG equation physically sensible, 
we transform variables to be dimensionless, where
the dimensional unit is taken to be the cutoff at that `time'.
This transformation runs the coupling constants, 
which is called the canonical scaling.
Also we introduce the wave function renormalization to normalize the
kinetic terms of every fields, which chooses a representative of
the equivalent classes in the theory space.
Then we finally get 
\begin{eqnarray*}
\frac{\partial\Se}{\partial t}&=
&-\frac{1}{2\delta t}\int'_p
\left\{
-\hbox{Tr}\ln
\left(
\frac{\delta^2 \Se}{\delta\phi_p\delta\phi_{-p}}
\right)
+\frac{\delta \Se}{\delta\phi_p}
\left(
\frac{\delta^2 \Se}{\delta\phi_p\delta\phi_{-p}}
\right)^{-1}
\frac{\delta \Se}{\delta\phi_{-p}}
\right\}\\
&&+d\Se+\int_p ~\phi_p\left(
\frac{2-d-\eta}{2}-p^{\mu}\frac{\partial'}{\partial p^{\mu}}
\right)
\frac{\delta}{\delta \phi_p}\Se\ ,
\end{eqnarray*}
where $d$ is the space-time dimension, $\eta$ is the anomalous dimension.
This equation, called the Wegner--Houghton equation,\cite{WH} describes 
the change of the effective action, that is, of a set of infinite number
of coupling constants. Thus the right-hand side
should be called a $\beta$ {\sl functional}, which controls the flows
in the infinite dimensional theory space.
Important is that it is obtained as an exact and closed form,
therefore it is sometimes called the exact renormalization group.


\section{Approximation method}

Even though we obtained the exact renormalization group, it is 
impossible to solve such an infinite dimensional functional differential
equation. Thus what is more important is if there is a good way of 
systematic approximation or not. The approximation method for 
the exact renormalization group is a projection to a sub-theory
space. For example, we take a sub-theory space spanned only
by the local potential terms in addition to the kinetic terms
$$
S_{\rm eff}=\int d^dx\left\{
V(\phi)+\frac{1}{2}(\partial_\mu\phi)^2
\right\}\ ,
$$
which is regarded as the lowest order of the expansion of the effective 
action with respect to the number of field derivatives.
This is called the local potential approximation 
(LPA),\cite{WH,lpa} where the
corrections to the kinetic terms are ignored either, thus it
gives vanishing anomalous dimensions.
The Wegner--Houghton equation in LPA is reduced to read
$$
\frac{\partial V(\phi, t)}{\partial t}
=\frac{A_d}{2}\left[{\rm ln}(1+V^{\prime\prime})\right]
+d{\cdot}V+{2-d\over 2}{\phi}V^\prime\ ,
$$
where $V^\prime$ denotes the derivative with respect to the
field $\phi$, and $A_d$ is a constant of the $d$-dimensional
angular integral.
We see the original functional differential equation is 
simplified to a 2-dimensional ($t,\phi$) partial differential
equation, which is non-linear, and is to be solved as
an initial value problem.
Note that LPA still deals with an infinite dimensional theory
space spanned by an arbitrary function $V(\phi)$.

It is still not easy to analyze the system precisely. 
We reduce it to a finite dimensional sub-theory space by expanding
the potential in terms of the polynomials in $\phi$. 
For example, take a 
scalar theory with Z$_2$ symmetry in 3 space-time dimension, and
expand the potential as
$$
V(\phi,t)=a_1(t)\rho + a_2(t)\rho^2 + a_3(t)\rho^3 \cdots\ \ ,\ \ \ \ 
\rho\equiv\phi^2/2\ .
$$
Then the $\beta$ functions (flow equations for the
coefficient functions) are given by 
\def\Ma{1+a_1}
\begin{eqnarray*}
\dot a_1 &=& {\Green 2a_1} + {3\over 2\pi^2}\ {a_2 \over {\Red \Ma}}, 
\hskip75mm
\\
\dot a_2 &=& {\Green a_2}  - {9\over 2\pi^2}\ {a_2^2\over {\Red (\Ma)^2}}
            + {5\over \pi^2}\ {a_3 \over {\Red \Ma}}, \\
\dot a_3 &=& {27 \over 2\pi^2}\ {a_2^3 \over {\Red (\Ma)^3}}
            - {45 \over 2\pi^2}\ {a_2 a_3 \over {\Red (\Ma)^2}}.
\end{eqnarray*}
\vskip-30mm
\hbox to \hsize{\hfil\epsfxsize=47mm\epsfbox{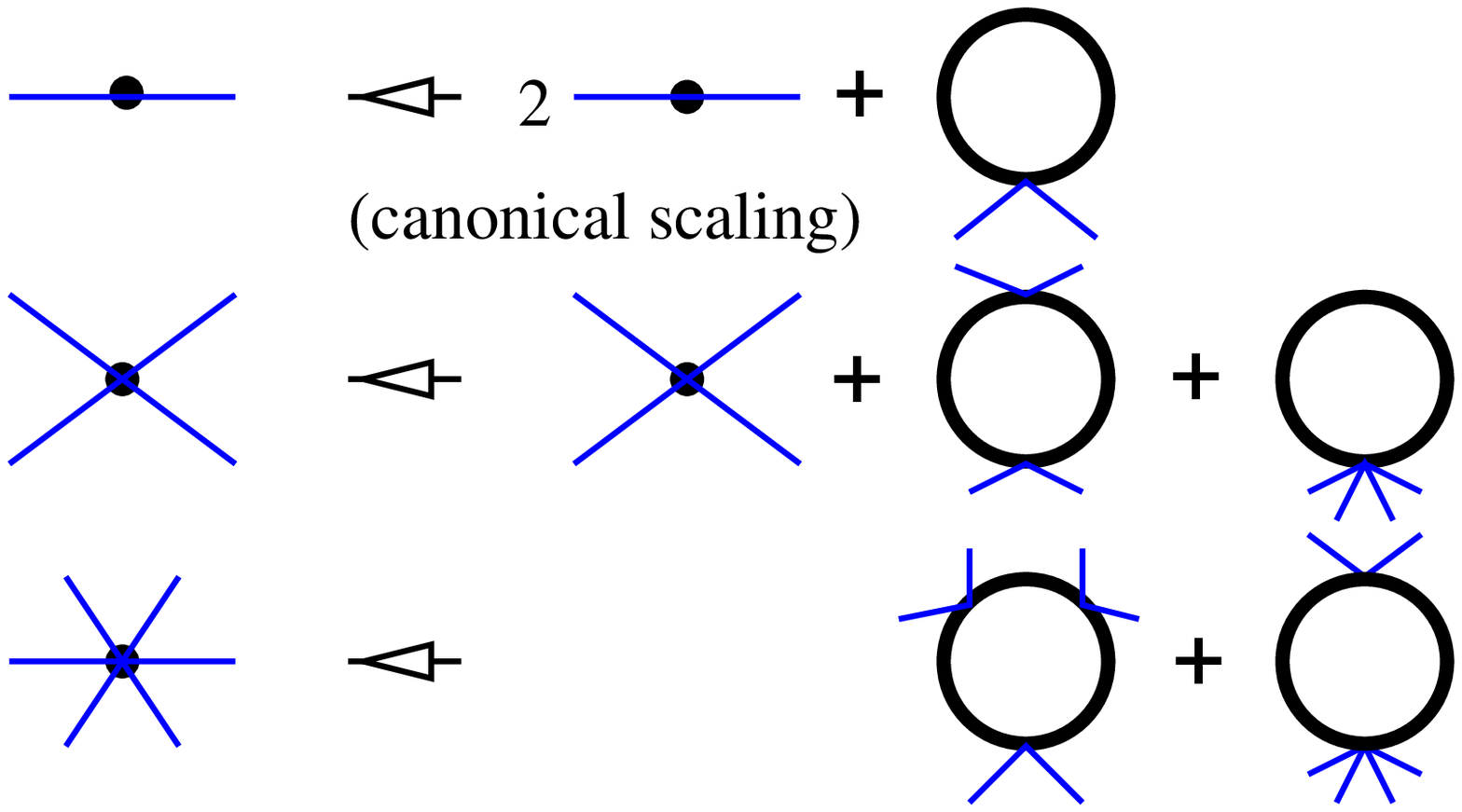}\hskip2mm}
\vskip2mm
\noindent
Even in this crude approximation with only 3 operators, we find a 
non-trivial fixed point realizing the ferromagnetic phase transition
with the critical exponent $\nu = 0.586$, where the lattice simulation
gives 0.629, such a good coincidence.
We increase the number of operators or the
dimension of the sub-theory space to see if the results 
converge. 
If it does, we understand we get the results for LPA sub-theory space.

\hangindent-59mm
\hangafter3
The approximation here is a projection to a sub-theory space, 
giving a projected $\beta$ functions, which define
a projected flow. The projected flow is different in general from
the projection of the true flow. The true flow always goes out of the
sub-theory space. The projected flow pulls it back on the sub-theory 
space in each step of transformation. In some exceptional case
as the large-$N$ theory, the projected flow coincides with the
projection of the true flow if we choose a good coordinate system
of the theory space, which we call the perfect 
coordinate.\cite{lpawh}

\vskip-34mm
\hbox to \hsize{\hfil\epsfxsize=54mm\epsfbox{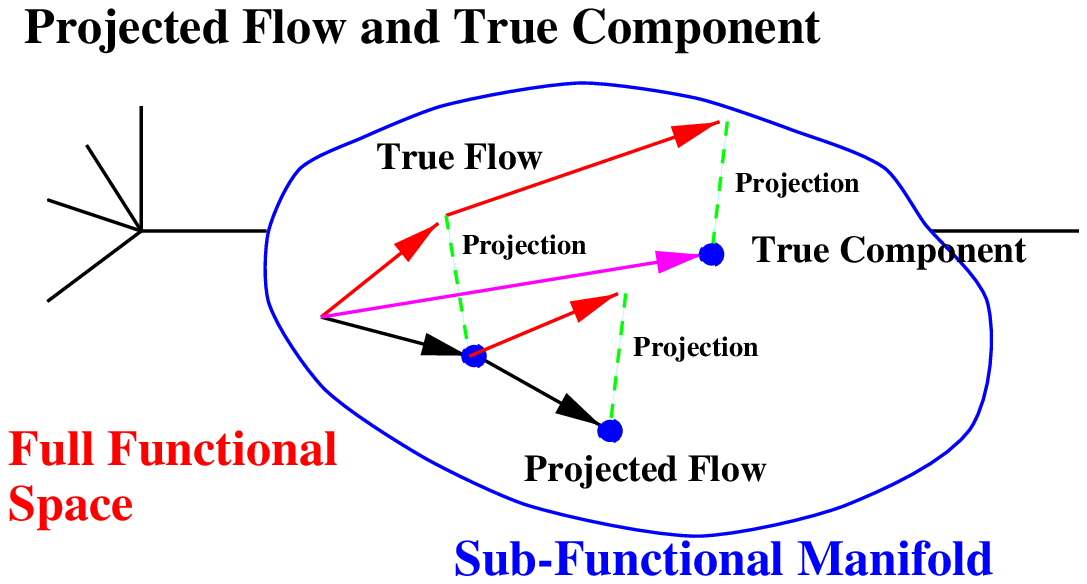}\hskip2mm}
\vskip4mm
Generally we may  expect that enlarging the sub-theory space,
the physical results converge, or oscillate at worst. 
This is a natural expectation.
Enlarging the sub-theory space is not directly related to any divergent 
series expansion. It is more like enlarging the total lattice
size in simulation, where nobody expects the results to diverge.
However, usually we must think out a good
series of sub-theory spaces to get fast convergence.
We have learned that an `environmentally friendly'
coordinate system assures better convergence. For example, 
expanding the potential, we might define
the following two schemes:
a fixed scheme (A) and a comoving scheme (B),\cite{comoving}
$$
\hbox{A}\Black :\ \ V(\rho)=\sum_{n=1}^{nmax} a_n \rho^n \hskip20mm
\hbox{B}\Black :\ \ V(\rho)=\sum_{n=1}^{nmax} a_n (\rho - \rho_0(t))^n
$$
The comoving scheme is a dynamic coordinate since its 
origin of expansion $\rho_0(t)$ is taken as a minimum of the potential at 
time $t$. Physically it is better since it uses more appropriate
propagators at $t$, and it actually improves convergence
drastically as is seen in Fig.\ref{fig-comoving}.\cite{lpawh,SCHEME,Ascheme}
\footnote{After the conference, we find that even the comoving frame
gives finally diverging results. 
However, it is at extremely large orders, thus it is after the highly
accurate convergence of even 8 digits for $\nu$ 
has been obtained\cite{Ascheme}.
We thank T.~R.~Morris to motivate us to investigate very large
orders.
}


\section{Comparison with other non-perturbative methods}

We compare our NPRG results
with other standard non-perturbative methods.
First let us examine the leading critical exponent of 
$d$-dimensional scalar theory comparing with the
$\epsilon$-expansion.\cite{lpawh}
As is seen in Fig.\ref{fig-epsilon}, 
the $\epsilon$-expansion gives a divergent series. Of course it 
is extensively investigated that these divergent series can be
Borel resummed to give a convergent value which is also 
displayed in the figure. Our NPRG analysis gives a globally
right behavior in the lowest order approximation LPA.

\vskip3mm
\refstepcounter{figure}
\label{fig-comoving}
\refstepcounter{figure}
\label{fig-epsilon}

\centerline{\hfil\epsfxsize=60mm\epsfbox{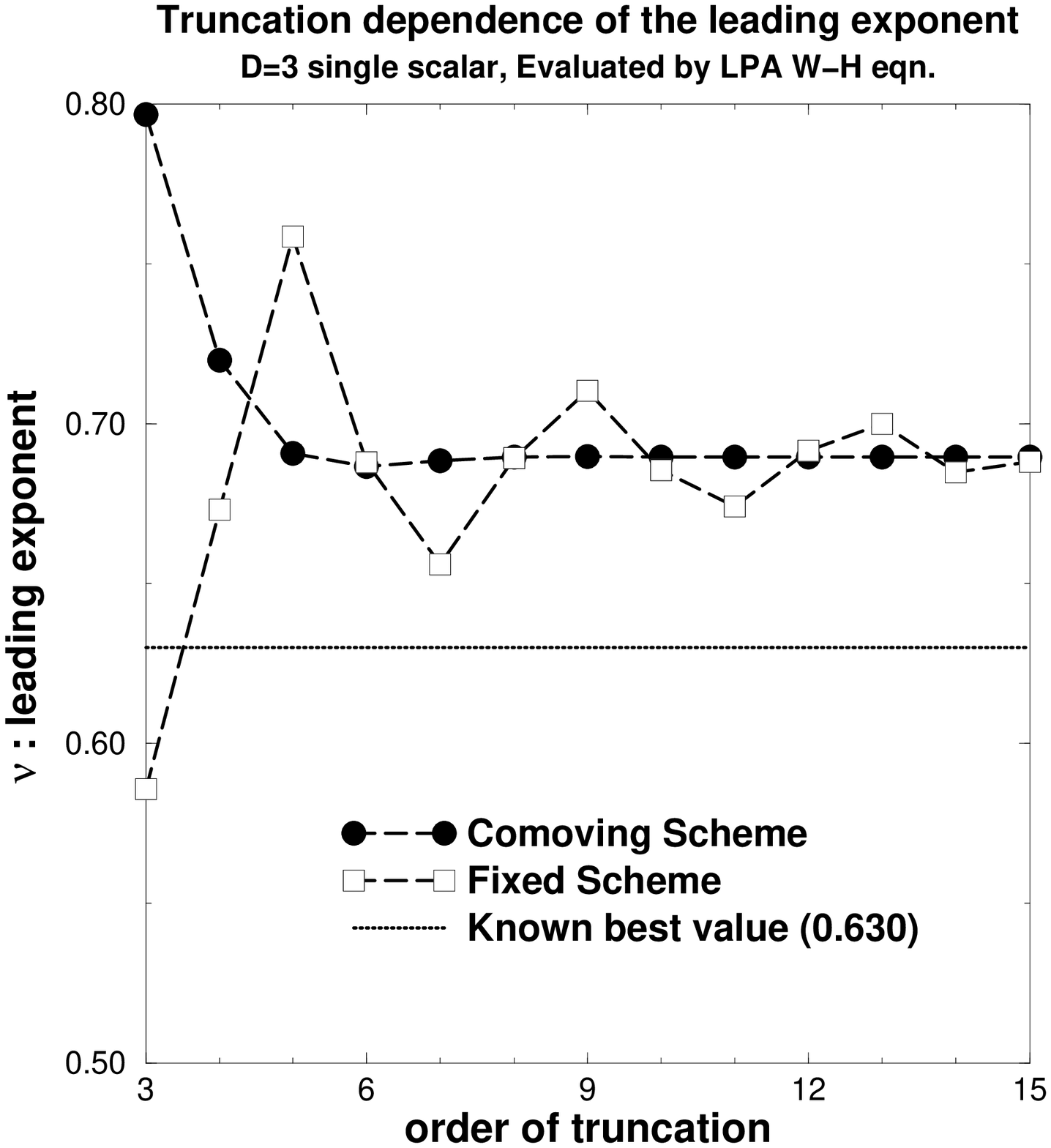}
\hfil\epsfxsize=60mm\epsfbox{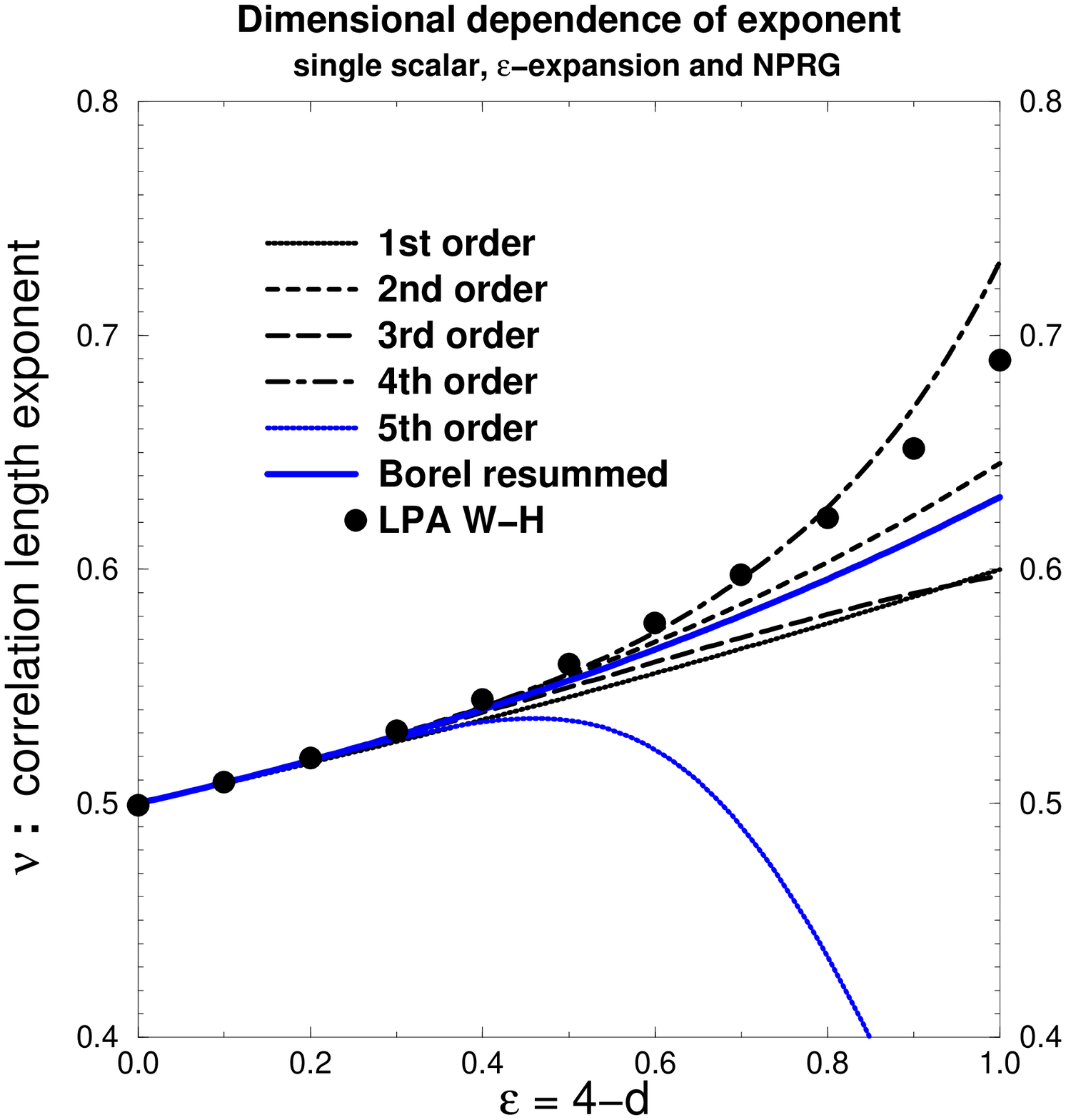}\hfil}

\vskip-6mm
\centerline{\Figfont\hfil Fig.\ref{fig-comoving}\hfil\hfil
 \hskip10mmFig.\ref{fig-epsilon}\hfil }
\vskip1mm

Next we analyze scalar theories with $N$-components
in 3 space-time dimension,  and
compare NPRG with the $1/N$ expansion 
(Fig.\ref{fig-largeN}).\cite{lpawh,souma-pw,morris-de}
The $1/N$ leading
result is just $\nu=1$, and we show the next to leading (O($1/N$))
and the next next to leading  (O($1/N^2$)) results. 
It is clearly seen that the $1/N$ expansion is also a divergent series. 
For small $N$ theories, the leading result ($\nu=1$ classical
result) is the best, and adding higher orders in $1/N$ makes it worse.
Our NPRG analysis gives a stable and globally correct results also in 
this comparison, and becomes exact at $N \rightarrow \infty$ limit.
We expect that when enlarging the sub-theory space
beyond LPA, we get closer to the true values.
Some of the beyond LPA results are also plotted in Fig.\ref{fig-largeN}
and our expectation proves right at least in the small $N$ region.
Note that Borel resummation for $1/N$ expansion is
not feasible at all. 

\refstepcounter{figure}
\label{fig-largeN}
\refstepcounter{figure}
\label{fig-comp-largeN}

\vskip3mm
\centerline{\hfil\epsfxsize=65mm\epsfbox{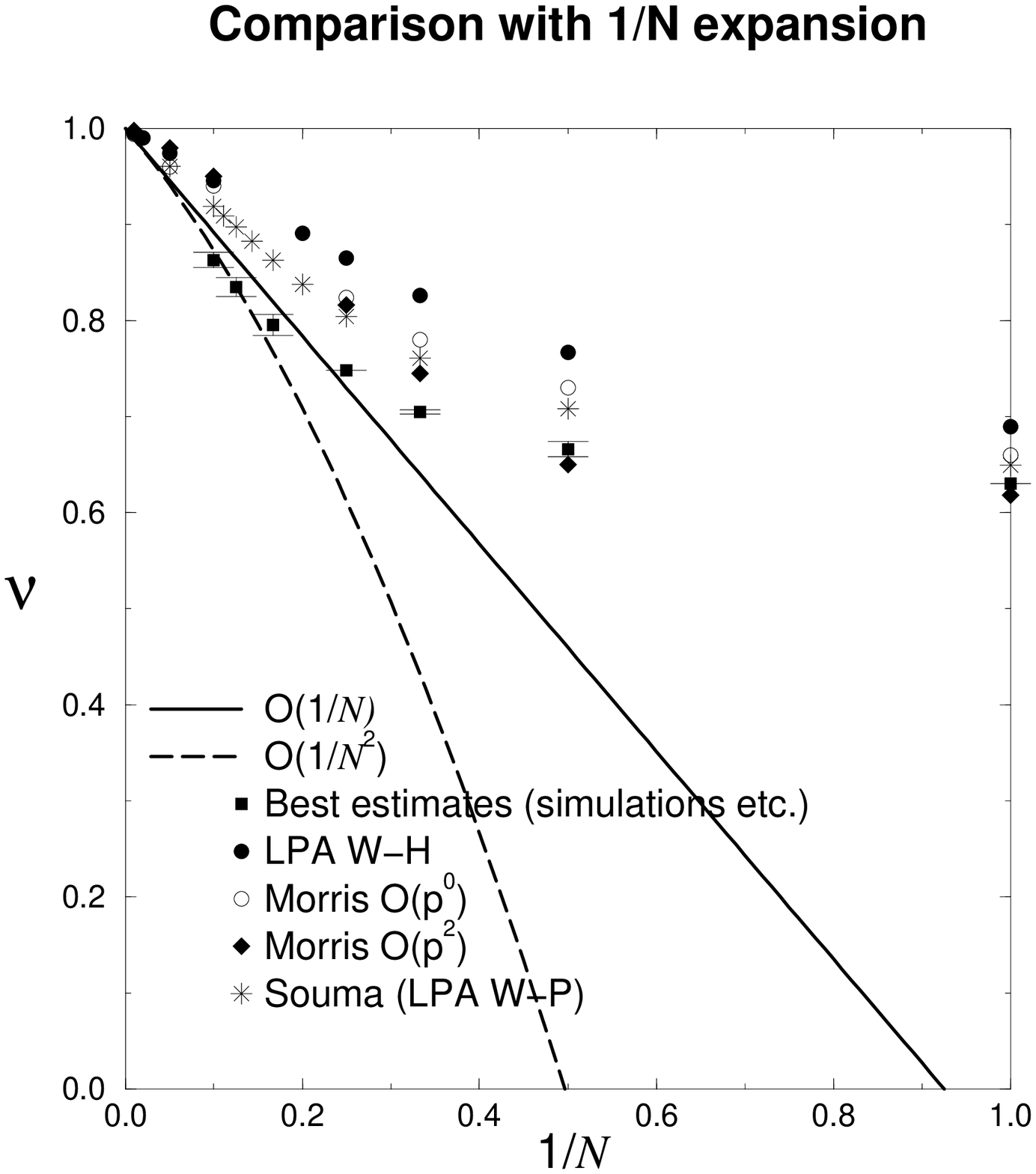}
\hfil\vbox{\hsize=60mm\epsfxsize=50mm\epsfbox{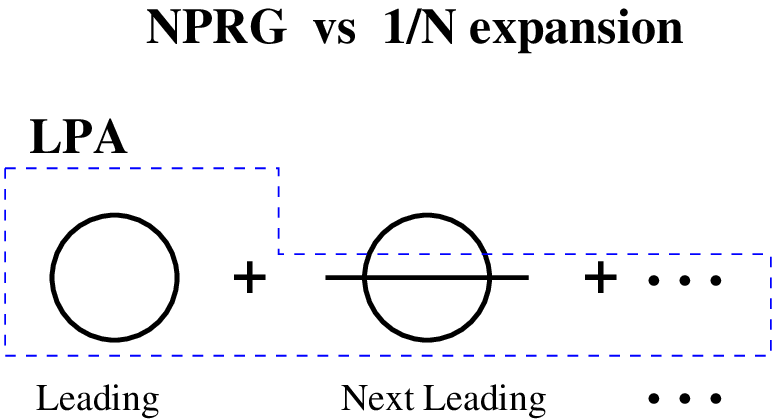}\vskip11mm}\hfil
\hskip5mm}

\vskip-7mm
\centerline{\Figfont\hfil\hbox to 65mm{\hfil Fig.\ref{fig-largeN}\hfil}\hfil
\hbox to 50mm{\hfil Fig.\ref{fig-comp-largeN}\hfil}\hfil}
\vskip3mm

What is a crucial difference between NPRG and, for example, 
$1/N$ expansion? 
The $1/N$ expansion contains only limited structures of diagrams
corresponding to the order of the expansion, while NPRG contains
every structured diagrams, exactly for the leading, 
but not exactly for higher orders in $1/N$, even at the lowest level of
LPA. 
Generally NPRG contains every powers of parameters as
$\epsilon$, $1/N$, and coupling constants, and
it contains the lowest order exactly. This feature assures that
NPRG gives globally correct behavior, and it is expected to converge when
enlarging the sub-theory space.

We show another striking results in case of 
scalar theories in 1 space-time dimension
also known as the unharmonic oscillator,\cite{tunnel}
whose action is given by
$$
S = \int dt\ \  \left({1\over 2} \dot x^2 \pm {1\over 2}x^2 + \lambda x^4
    \right)\ .
$$
We evaluate the energy gap between the ground state and the first
excited state using the infrared effective potential obtained by
NPRG. 
In case of the single-well potential (Fig.\ref{fig-single}),
our results are almost exact for whole range of the coupling 
constant $\lambda$, whereas the standard perturbation series
shows its divergent nature even at the weak coupling region
(although Borel resummation is possible in this case).

\refstepcounter{figure}
\label{fig-single}
\refstepcounter{figure}
\label{fig-double}

\vskip2mm
\centerline{\hfil\epsfxsize=50mm\epsfbox{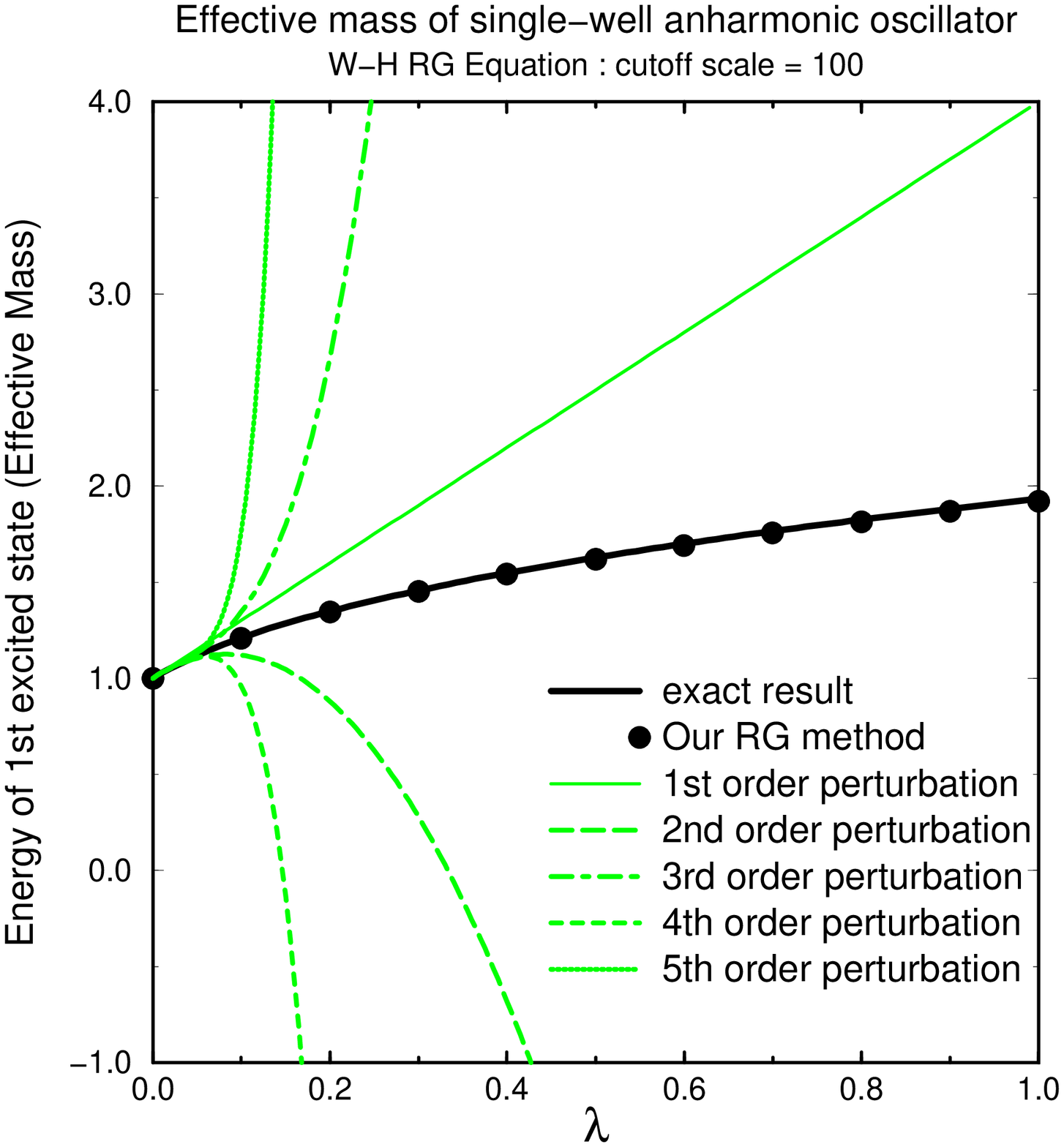}
\hfil\epsfxsize=50mm\epsfbox{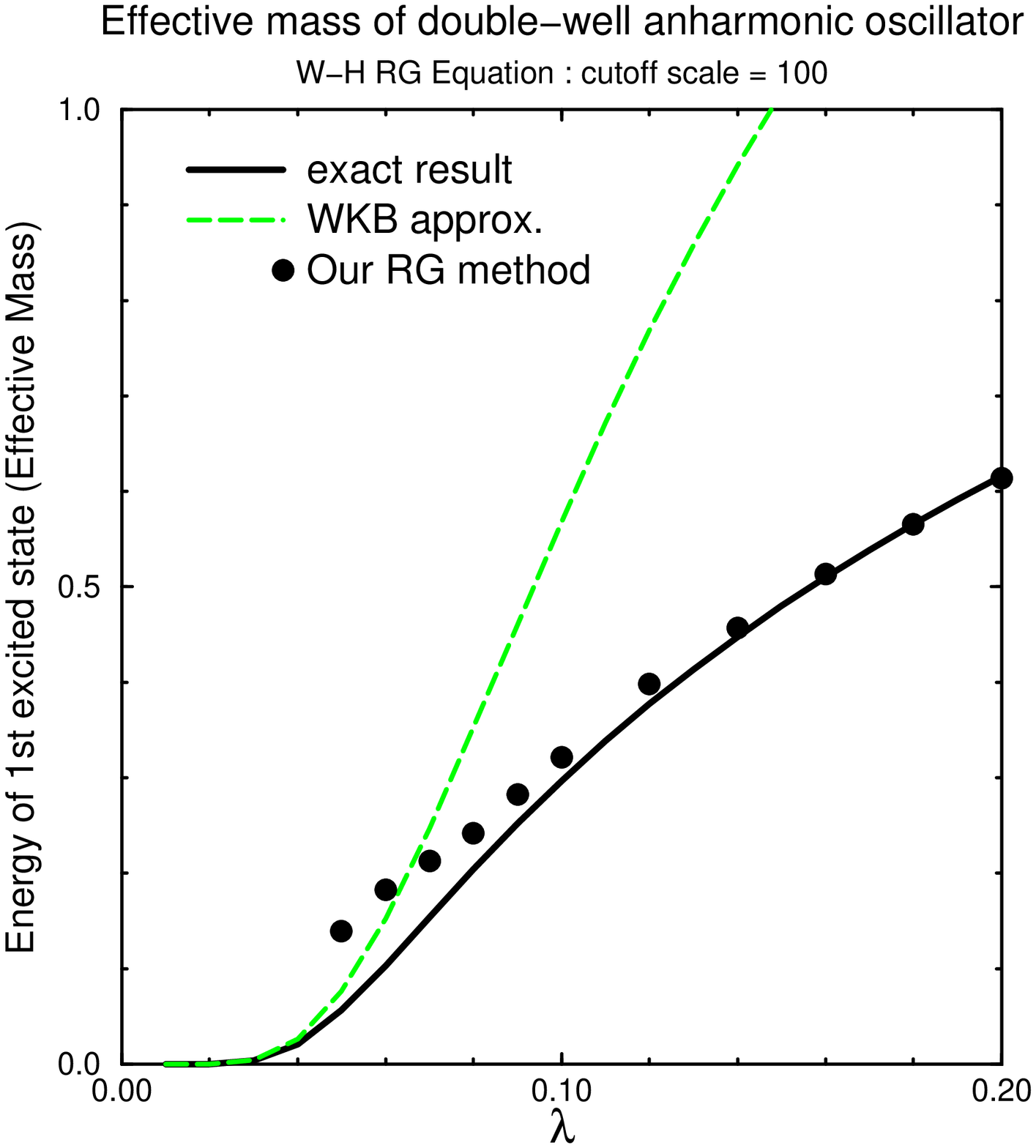}\hfil}

\vskip-5mm
\centerline{\Figfont\hfil\hbox to 50mm{\hfil Fig.\ref{fig-single}\hfil}\hfil
\hbox to 50mm{\hfil Fig.\ref{fig-double}\hfil}\hfil}
\vskip2mm
\noindent
In case of the double-well potential, we compare our results with
the instanton calculation (Fig.\ref{fig-double}). At $\lambda\sim 0.08$
the first excited state
energy crosses down the central barrier, where the LPA results
start deviating from the exact values. 
At this point the instanton results are already unreliable. 
The LPA works extremely well except for the deep region 
where the tunneling rate is very small and the dilute instantons 
become exact.

To see how NPRG works well for 
the double-well potential is very important, since it is related to
a more general issue of how NPRG can treat essential
singularities and topological objects in field theory.
The LPA gives a convex effective potential in case of the spontaneous
symmetry breaking, which indicates that LPA is able to take
account of the global domain wall structures without which the
convexity may not appear.\cite{tunnel}
However it seems that LPA 
is not successful to evaluate precisely the action accompanying such
topological configurations since it fails to give the correct energy gap 
in case of low tunneling rate. 

\section{Non-perturbative RGE vs. Perturbative RGE}

Here we recapitulate the features of NPRG, compared to the standard
perturbative RGE (PRGE). PRGE defines the $\beta$ functions which are 
calculated by the loop expansion, giving the leading log series
expansion or the improved perturbation, while NPRG $\beta$ functional
is infinite dimensional, but it is exactly given by tree and one-loop
diagrams. Multi-loop effects come out through the higher dimensional
operators. PRGE gives the asymptotic series at best, and actually
other non-perturbative methods including the $\epsilon$-expansion and
the $1/N$-expansion do as well, while in NPRG the approximation is 
defined by a projection to a sub-theory space and systematic
enlargement of it, thus no diverging series may
appear since no expansion in any parameter is employed.

The standard effective action $\Gamma(\phi)$ obtained by PRGE
does not serve 
the dynamical symmetry breaking due to a composite order parameter.
We need to evaluate Cornwall-Jackiw-Tomboulis (CJT) effective 
potential for the composite
operators or equivalently to solve the Schwinger-Dyson (SD)
equation.\cite{CJT}
However, NPRG automatically contains the CJT effective potential
through the Wilsonian
effective action including higher dimensional multi composite
operators, thus the non-perturbative vacuum can be directly analyzed. 

NPRG cannot manifest the gauge invariance due to the momentum
cutoff. Then theory space must span gauge non-invariant operators 
as well, and we have to pick up the gauge invariant theory space
(which differs from the gauge invariant 
operator space).\cite{QEDscheme,MSTI}
This is a serious problem for load of the practical calculation, 
but it is not an essential difficulty and can be treated
anyway.

To summarize, NPRG method is very general, robust and global, 
and no divergent series appears. Recent development has been done
by numerical integration of the NPRG equation in LPA and
beyond LPA sub-theory spaces. This is a balanced research
between analytic and numeric. The lowest order sub-theory space, 
LPA, has exhibited the features that it contains one-loop perturbative
results, exact in the large $N$ leading, exact in the 1st order
of $\epsilon$-expansion, and furthermore contains the improved ladder SD
results.


\section{Dynamical chiral symmetry breaking}

Due to shortage of pages I have, we just briefly describe the basic
results of our works.\cite{chiral}
We consider a chiral invariant gauge theory and set up NPRG
equations in LPA. We include multi-fermi operators, 
4-fermi, 8-fermi, $\cdots$, respecting the chiral 
invariance. The $\beta$ functions for 4-fermi operators do not
depend on 8- or more-fermi operators due to the chiral invariance.
Thus our multi-fermi coordinates is a perfect coordinates and
the flow of 4-fermi operators can be solved exactly in LPA, 
coupled with the 
running of the gauge coupling constant $e$ ($\lambda = 3e^2/4\pi^2$):
\begin{eqnarray*}
{d \GS\over dt} &=& -2 \GS + \left(2 - 
{\Red {1\over2}}\right) \GS^2 - {\Red 4 \GS \GV}
+ \lambda \GS  + \left({1\over 8} + {\Red {1 \over 24}}\right)
  \lambda^2\ ,\\
\noalign{\vskip3mm}
{\Red {d \GV\over dt}} &=& {\Red -2 \GV + {1\over4}\GS^2 
          - \lambda \GV - {1\over 12} \lambda^2 }\ .
\end{eqnarray*}
Figure \ref{fig-4f} shows the diagrams contributing to the 
$\beta$ functions of the scalar 4-fermi $\GS$ 
and vector 4-fermi $\GV$ interactions, where the
dashed lines represent the gauge boson.

\refstepcounter{figure}
\label{fig-4f}
\refstepcounter{figure}
\label{fig-4fbeta}
\vskip4mm
\centerline{\hfil\epsfxsize=48mm\epsfbox{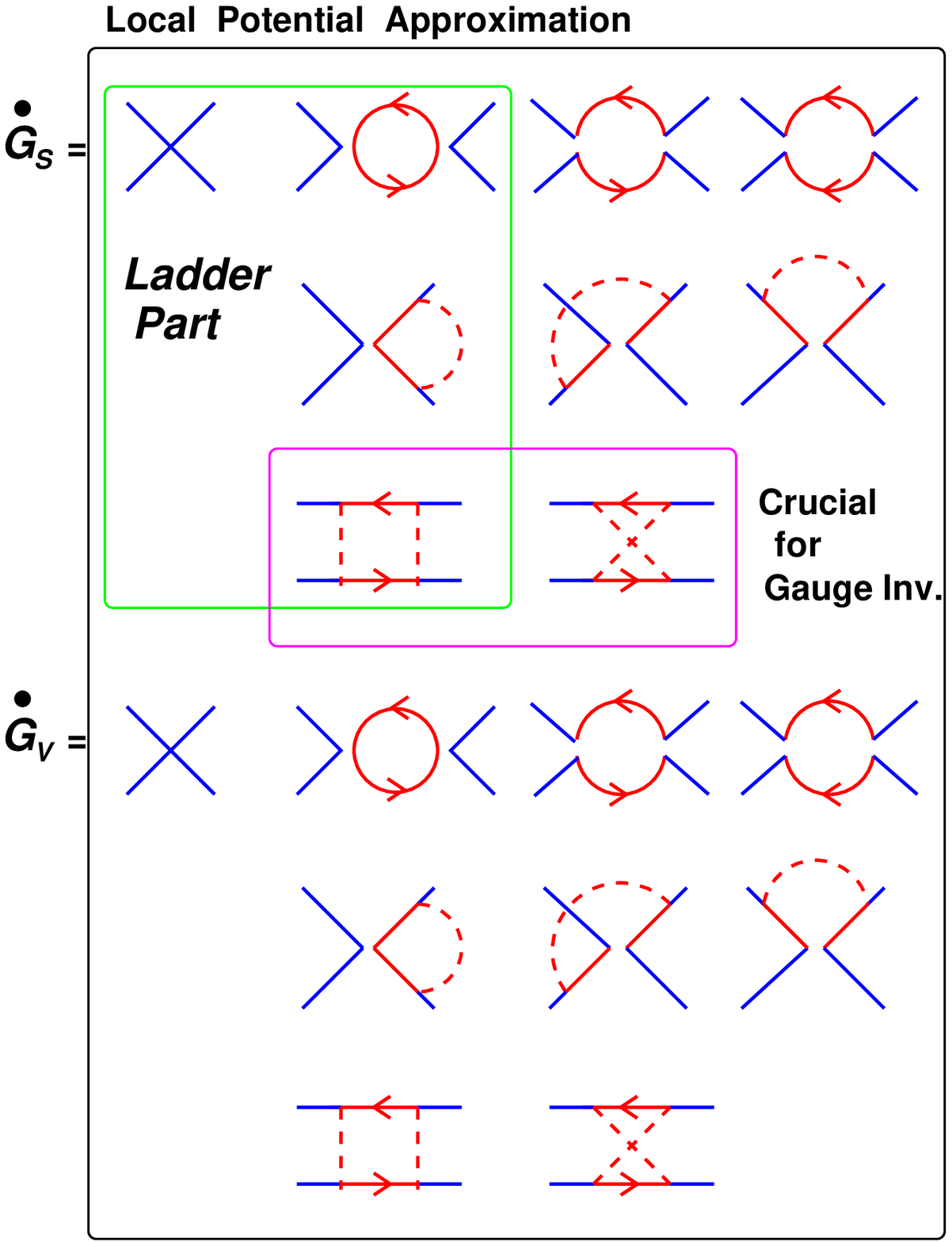}
\hfil\hfil\epsfxsize=65mm\epsfbox{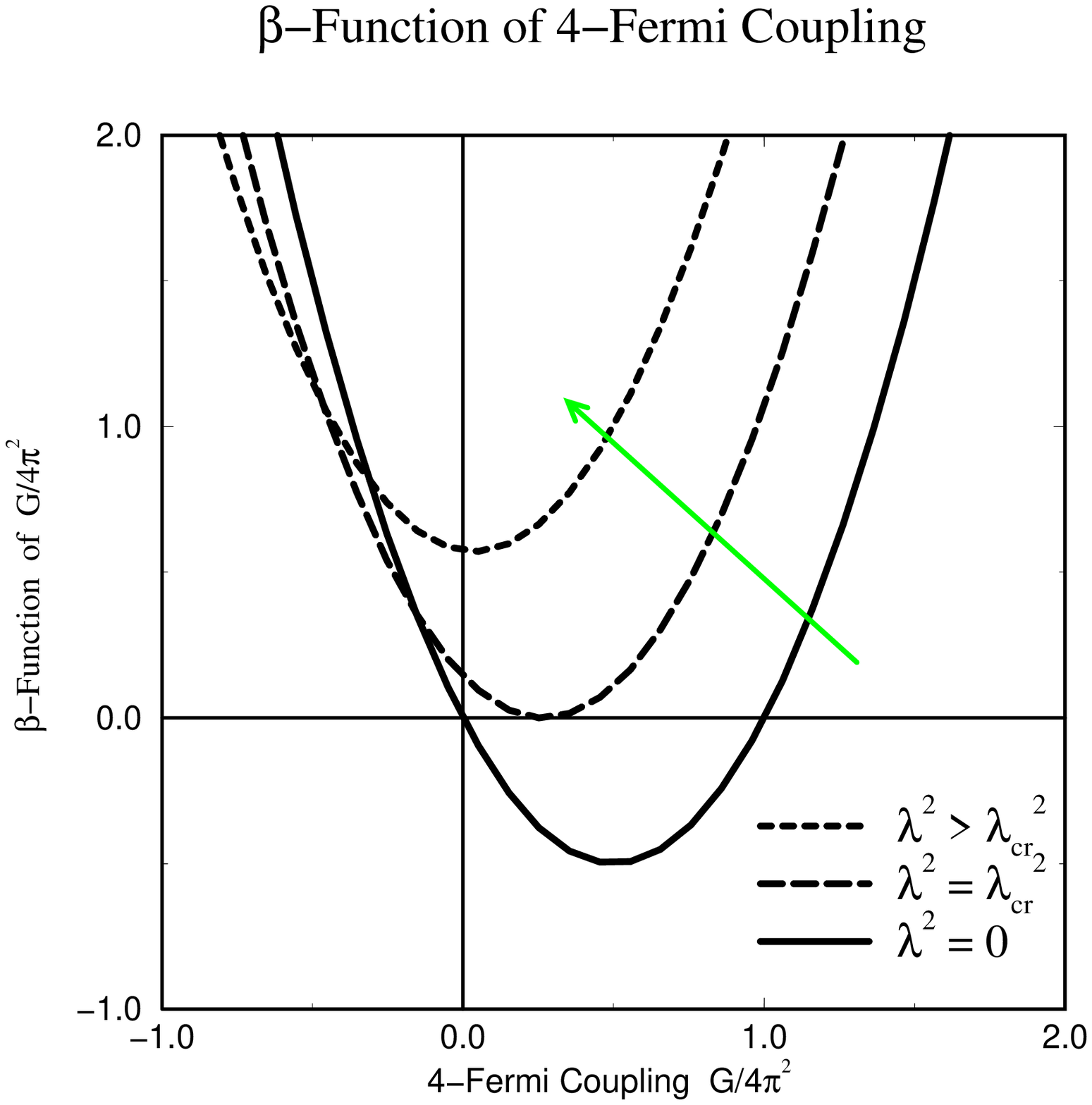}\hfil}

\vskip-0mm
\centerline{\Figfont\hfil\hbox to 48mm{\hfil Fig.\ref{fig-4f}\hfil}\hfil\hfil
\hbox to 65mm{\hfil Fig.\ref{fig-4fbeta}\hfil}\hfil }
\vskip1mm

We define a part of the $\beta$ function, the ladder part, as in
Fig.\ref{fig-4f}. 
If one takes only this part in the following
analyses, one obtains exactly the same results as those by the ladder
SD equation in the Landau gauge. 
(This uniqueness of  
the Landau gauge comes from the fact that the fermion anomalous
dimension vanishes in this gauge.) 
Remember that the ladder SD suffers 
a strong gauge dependence,\cite{SDdifficulty} whose
reason is clear here. The ladder picks up just a part of the
gauge invariant set of diagrams in the $\beta$ function, 
taking the ladder but not the 
crossed ladder. On the other hand, NPRG $\beta$ function takes both, 
no difference for either ladder or crossed ladder, thus it respects 
the gauge independence of flows and of the physical results. 

One readily sees (Fig.\ref{fig-4fbeta}) that at $\lambda=0$, 
the scalar 4-fermi flows has an ultraviolet
fixed point which is nothing but the Nambu-Jona-Lasinio (NJL) criticality,
and an infrared fixed point of the Gaussian one.
There appear two phases, strong and weak.
Note that the 8-fermi operator has no fixed point without gauge
interactions, indicating the triviality of the pure fermi theory.
When the gauge coupling is switched on, the Gaussian fixed point moves
up, the NJL fixed point moves down, and they finally meet together
to pair annihilate at a critical gauge coupling $\lambda_{\mbox{c}}$,
above which there is only one phase left.
To identify these two phases, we investigate the CJT effective
potential for the chiral condensate $<\bar\psi\psi>$, and find that
the weak phase is a symmetric phase, while the strong phase exhibits
the spontaneous symmetry breaking of the chiral symmetry, where the
potential takes the flat bottomed convex shape characteristic to the
spontaneous symmetry breakdown. 
This structure is displayed as a phase diagram in the scalar 4-fermi 
vs. the gauge coupling plane in Fig.\ref{fig-phase}, which should be
compared with the old traditional phase diagram.\cite{SDfourfermi}
We may calculate a physical quantity of this system, the anomalous
dimension of the mass operator $\bar\psi\psi$ (Fig.\ref{fig-gamma}). 
It is enhanced compared to the ladder SD results. Note that
our results are LPA exact, including `beyond the ladder' diagrams, and
are gauge independent.

\refstepcounter{figure}
\label{fig-phase}
\refstepcounter{figure}
\label{fig-gamma}
\vskip2mm
\centerline{\hfil\epsfxsize=60mm\epsfbox{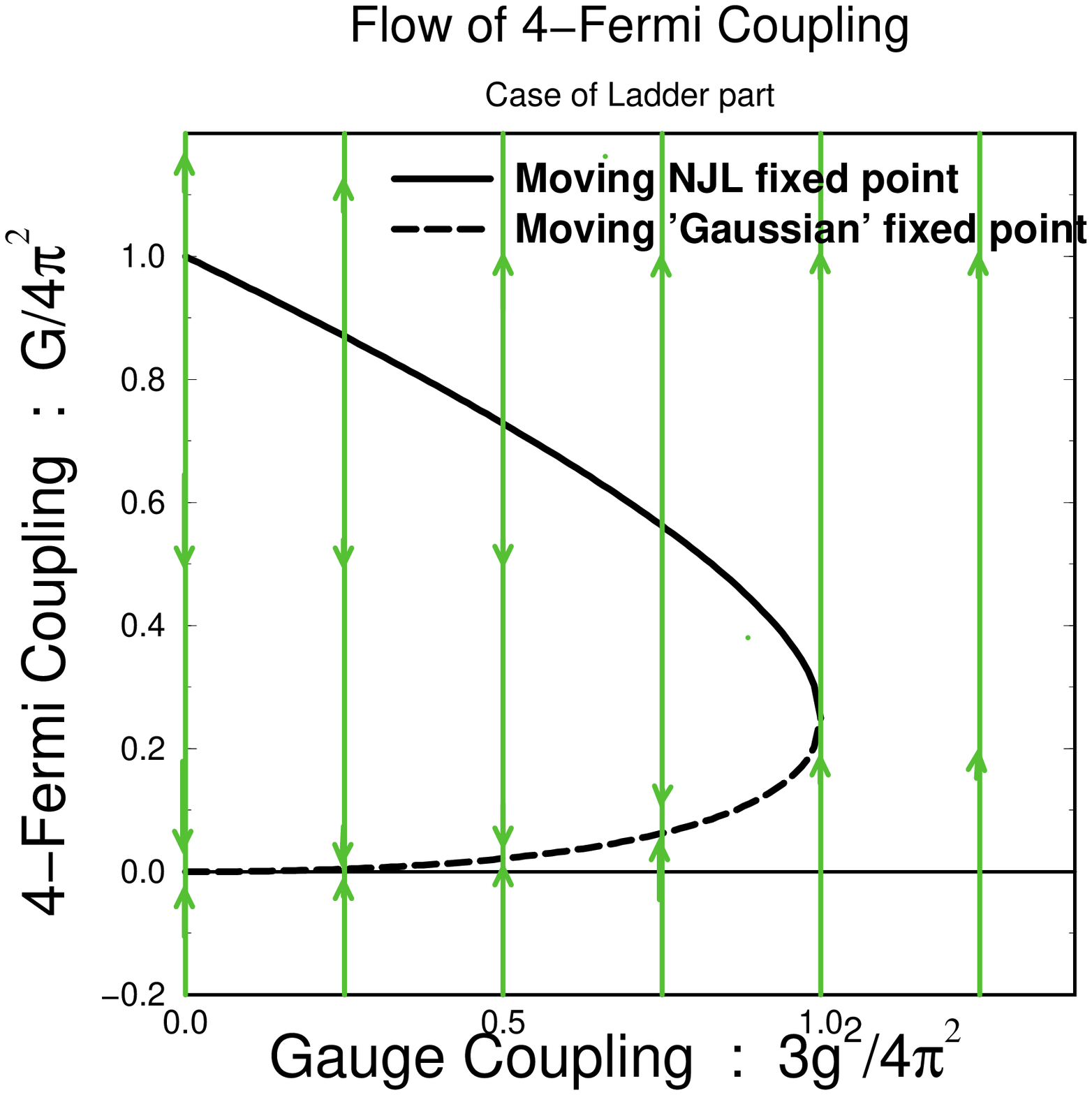}
\hfil\epsfxsize=60mm\epsfbox{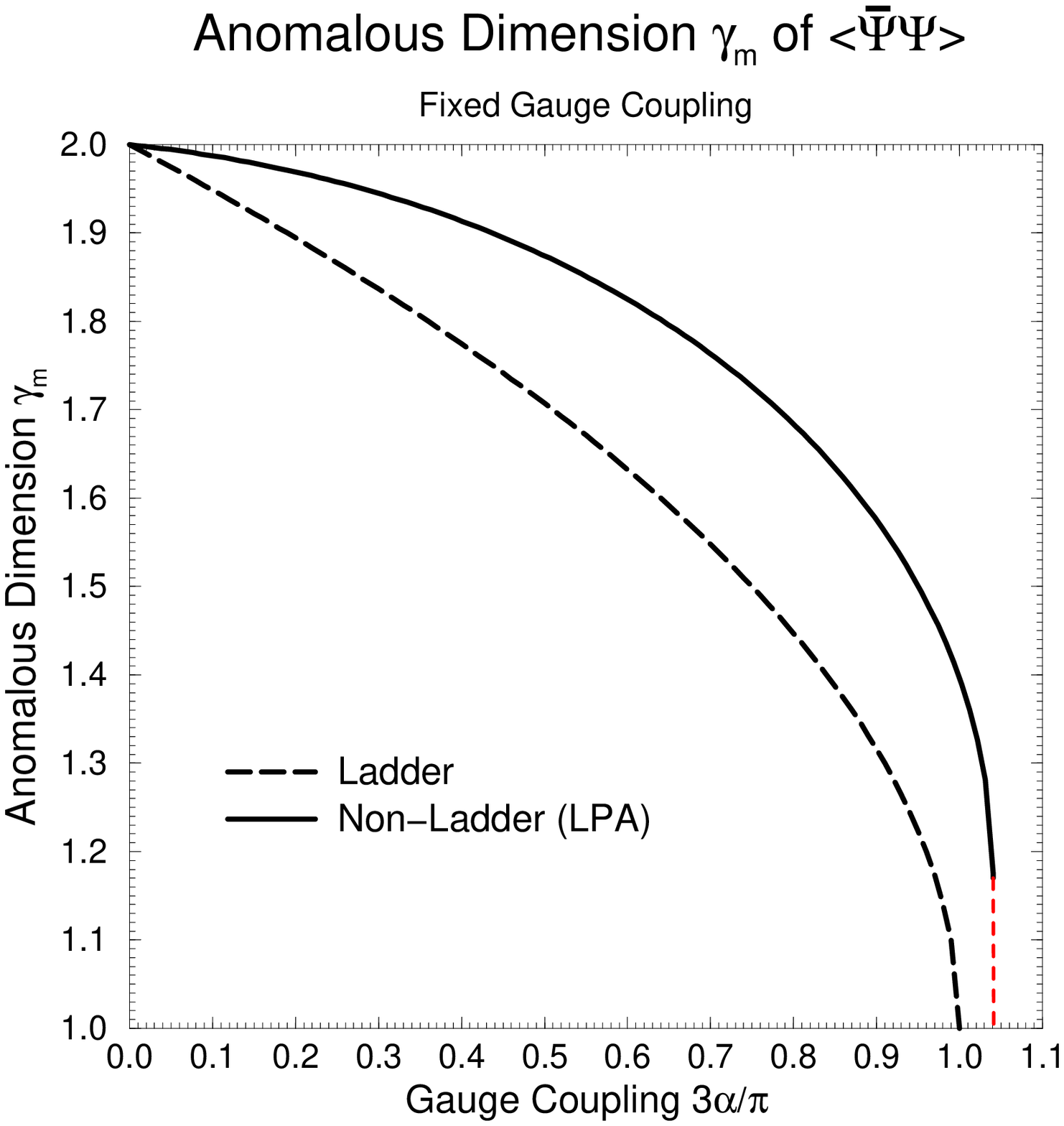}\hfil}

\vskip-0mm
\centerline{\Figfont\hfil\hbox to 60mm{\hfil Fig.\ref{fig-phase}\hfil}\hfil
\hbox to 60mm{\hfil  Fig.\ref{fig-gamma}\hfil}\hfil }

We may run the gauge coupling constant. The U(1) gauge interactions
keep the two phase structure. However the fixed points are
NJL and the Gaussian only, and the renormalized trajectory is on the
trivial line, which is consistent with 
the recent lattice simulations.\cite{latticeQED}
Asymptotically free gauge interactions wipe out the symmetric phase, 
and only the Gaussian fixed point survives.
Therefore the whole region of the theory space belongs to 
the chiral
symmetry broken phase. We can evaluate physical quantities, 
the chiral condensate and the fermion mass.
In this running gauge coupling case, we proved that the ladder 
part $\beta$ function now gives exactly the same results as those
by the improved ladder SD equation in the 
Landau gauge,  {\sl a la} Higashijima,\cite{Higashijima} where
the scale parameter of the running gauge coupling constant takes
a larger momentum of the fermion legs. 
This particular definition of the
running gauge coupling constant exactly
emerges from the NPRG LPA with the ladder
part. Now we know what is included and what is missing in the
improved ladder SD equations, thus establishing the improved
ladder for the first time.

To evaluate the infrared physical quantities, the bare 
multi-fermi coordinate system is not good, since it uses the 
massless fermion propagators. We introduce a collective coordinate
of a scalar field coupled to the $\bar\psi\psi$ operator, which is 
another example of the environmentally friendly coordinate system.
Within the ladder part, we have calculated the chiral condensate
and the fermion mass as in Fig.\ref{fig-qcdres}, which shows good
convergence within low dimensional sub-theory spaces, and 
agrees with the improved ladder SD results.\cite{SDimproved}
Taking the non-ladder $\beta$ components into account, 
we proceed to evaluate
infrared physical quantities beyond the ladder
SD results, in a gauge independent way. That will be the first
results of gauge independent and including `beyond the ladder',
thought it is not an easy task.

\refstepcounter{figure}
\label{fig-qcdres}
\vskip2mm
\hbox to \hsize{\hfil\epsfxsize=69mm\epsfbox{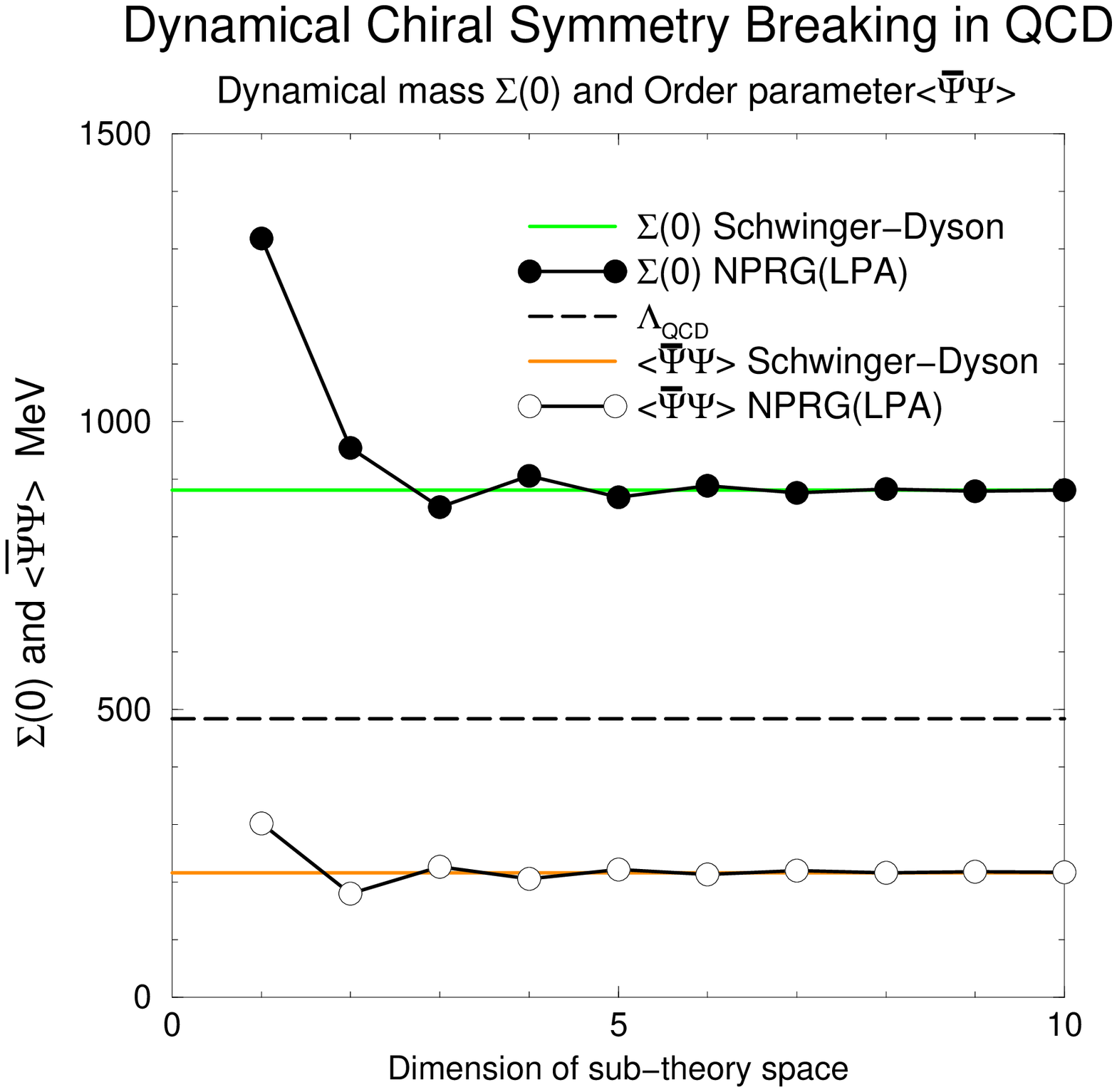}\hfil}
\hbox to \hsize{\Figfont
\hfil\hbox to 69mm{\hfil Fig.\ref{fig-qcdres}\hfil}\hfil}
\vskip1mm

To deal with the gauge interactions more seriously, we need to formulate 
NPRG with a smooth cutoff function.\cite{Polchinski,WETTERICH}
Then the cutoff scheme (the profile of the cutoff function)
dependence might cause a problem. However we find that indeed such
scheme dependence affects phase criticalities (the critical
coupling constants) which are themselves 
not physical quantities, but the physical quantities like the
anomalous dimension of the mass operator has only very 
small scheme dependence.\cite{QEDscheme}

We have clarified that NPRG
equipped with the sub-theory space approximation potentially 
solves the dynamical chiral symmetry breaking in gauge theories, 
which can go beyond the ladder SD results restoring the gauge
independence. This new method of treating non-perturbative QCD
will give us a promising tool alternative to the lattice simulations.
However we may need some revolutionary new techniques of evaluating the
$\beta$ functional to solve QCD with larger dimensional
sub-theory spaces of the gauge sector keeping the
gauge invariance.

\section*{Acknowledgments}

The author is partially supported
by Grant-in Aid for Scientific Research
(\#08240216 and \#08640361) from the Ministry of Education,
Science and Culture.

\setlength{\parskip}{1mm}
\setlength{\baselineskip}{4.2mm}
\setlength{\rightskip}{0pt plus 50mm}

\end{document}